\documentclass[twocolumn,showpacs, preprintnumbers,amsmath,amssymb]{revtex4}
\usepackage{graphicx}
\usepackage{dcolumn}
\usepackage{bm}
\begin{document}
\title{
``Bimaximal + Democratic" type neutrino mass matrix in view of 
 quark-lepton complementarity  
}
\author{\bf 
Ambar Ghosal} 
\email{ambar.ghosal@saha.ac.in}
\affiliation{
Saha Institute of Nuclear Physics,
1/AF Bidhannagar, Kolkata 700 064, India
}
\author{Debasish Majumdar 
}
\email{debasish.majumdar@saha.ac.in}
\affiliation{
Saha Institute of Nuclear Physics,
1/AF Bidhannagar, Kolkata 700 064, India
}
\date{\today}
\begin{abstract}
We demonstrate that `Bimaximal + Democratic' type neutrino 
mass matrix can accommodate the deviation of $\theta_\odot$ 
from its maximal value which referred in the literature 
as `quark-lepton complementarity' along with the other 
present day neutrino experimental results, namely, 
atmospheric, CHOOZ, neutrinoless double 
beta decay ($\beta\beta_{0\nu}$) and result obtained 
from WMAP experiment. We define a function 
$\chi_p$ in terms of solar and atmospheric neutrino mass
squared differences and solar neutrino mixing angle
(obtained from different experiments and our proposed 
texture). The masses and mixing angles are expressed 
in terms of three parameters in our proposed texture.
The allowed region of the texture parameters 
is obtained through minimization of  the above  function. 
The proposed texture 
crucially depends on the value of the experimental 
results of $\beta\beta_{0\nu}$ experiment among all
other above mentioned experiments. If, in future, 
$\beta\beta_{0\nu}$ experiments, namely, MOON, EXO, 
GENIUS shift the lower bound on $\langle m_{ee} \rangle$ 
at the higher side by one order, the present texture will 
be ruled out.   
\vspace{1pc}
\end{abstract}
\pacs{
PACS number(s): 13.38.Dg, 13.35.-r, 14.60 .-z, 14.60.Pq.}
\maketitle
\narrowtext
\section{
 Introduction}
The conjecture of neutrino flavor oscillation has been strengthened 
by the Super-Kamiokande (SK) \cite{SK} 
atmospheric neutrino experiment. This has 
also been substantiated by K2K long baseline neutrino experimental 
results \cite{k2k}. 
The best-fitted values obtained from SK experiment are  
given by  $\Delta m^2_{\rm{atm}}$ = $2.4\times {10}^{-3}$ $\rm{eV}^2$, 
$\sin^22\theta_{atm}$ = 1.0 \cite{skbest}.  
Moreover, global analysis of all solar neutrino experimental results 
, including SNO salt phase experiment, \cite{sno}  
is in favor of Large Mixing 
Angle (LMA) MSW solution of solar $\nu_e$ deficit problem and this 
has also in concordance with the KamLAND \cite{kam} reactor 
neutrino experimental results. Furthermore, the CHOOZ \cite{chooz} 
experiment has also put an upper bound on the $U_{e3}$ element 
of the lepton mixing matrix and combined analysis of solar, atmospheric 
and reactor data put bound on $U_{e3}$ as $|U_{e3}|\leq 0.22$ at 
$99.73\%$ c.l \cite{choozcom}.
There are two other experimental results, one of them, 
from WMAP experiment  \cite{wmap} 
on cosmic microwave background anisotropies gives an 
upper bound on the total neutrino mass as $\Sigma m_i\leq 0.63$ eV 
\cite{wmapres}. The second one, the neutrinoless double beta experiment 
\cite{2beta}
has put 
a bound on Majorana-type neutrino mass as $<m_{ee}>$ = (0.05 - 0.84) eV at 
$99\%$ c.l with an uncertanity of the nuclear matrix elements up to 
$50\%$. 
\par       
In view of the above it is worthwhile to investigate appropriate 
texture of lepton mass matrix which satisfies all those experimental
results. 
First of all, exact `bimaximal' texture is not admissible due to 
departure of $\theta_\odot$ from $\pi/4$ and the deviation is parametrised 
in terms of Cabibbo angle and 
the phenomena is referred as `quark-lepton complementarity' 
\cite{qlc}.
Keeping the charged lepton mass matrix flavor 
diagonal, in the present work, 
we consider a neutrino mass matrix of the form 
`bimaximal + democratic' where the `democratic' part of the neutrino 
mass matrix parametrise the difference $\pi/4 - \theta_\odot$. 
We have not addressed any typical 
model in this work, because, there are plenty of literature concerning 
bimaximal structure of neutrino 
mass matrix , however, we have indicated possible source of democratic 
type matrix explicit realization of which needs some extra flavor 
symmetry. 
The present texture gives rise to $\theta_{atm}$ maximal and vanishing 
value $\theta_{13}$.
We fit two neutrino mass-squared differences with the required 
values of solar and 
atmospheric neutrino experimental results and the remaining  mixing angle 
with 
the required values of solar neutrino neutrino mixing angle. 
These three parameters are fixed
by defining a function $\chi^2_p$ (see later) as
the sum of squares of the differences between the calculated values
of neutrino oscillation parameters (with the texture considered) and
the best fitted values of the same (obtained from different analyses)
and then minimising this function.
The plan of paper is as follows: In Section II, we propose the texture 
of the neutrino mass matrix and  its eigenvalues and mixing angles.
We discuss phenomenological fits of the 
model parameters through our defined $\chi^2_p$ function in Section III. 
Section IV contains our conclusion. 
  
\section{Proposed Texture}
We consider the following texture of the neutrino mass matrix 
$$
M_\nu = M_{bi} + M_{demo}
\eqno(2.1)
$$
where the first part $M_{bi}$ is given by 
$$
M_{bi} =\begin{pmatrix}0&a^\prime&a^\prime\cr
                         a^\prime&0&c^\prime\cr
                         a^\prime&c^\prime&0

\end{pmatrix}
\eqno(2.2)
$$
and the second part $M_{demo}$ is as 
$$
M_{demo} =d\begin{pmatrix}1&1&1\cr
                        1&1&1\cr
                        1&1&1
\end{pmatrix}
\eqno(2.3)
$$
where $d$ is some scale factor. Thus the entire mass matrix 
contains three parameters $a,c$ and $d$ 
and we consider all of them to be real. A possible source of 
the terms which gives rise to `democratic' type mass matrix is 
due to higher dimensional operator  
$$
{\cal L} =\frac{f_{ij}}{M}\displaystyle\sum_{i,j} 
l_{iL}l_{jL}\phi\phi
\eqno(2.4)
$$
through considering equality between the $f_{ij}$ couplings.
Diagonalizing the entire neutrino mass matrix 
we obtain
 the following mixing angles
$$
\theta_{23} = -\pi/4, \theta_{31} = 0
\eqno(2.5)
$$
$$
\tan^2\theta_{12} = \frac{d - m_1}{m_2 - d}
\eqno(2.6)
$$
and the eigenvalues are
$$
m_1 = {\frac{3d+c+x}{2}}
$$
$$
m_2 = {\frac{3d+c-x}{2}}
$$
$$
m_3 = -c
\eqno(2.7)
$$
where 
$$
x = \sqrt{{(c+d)}^2+ 8{(a+d)}^2}
\eqno(2.8)
$$ 
where $a = a^\prime + d$ and  $c = c^\prime + d$.
Next, we set the solar and
atmospheric neutrino mass squared
differences as
$$
\Delta m_{sol}^2 = \Delta m_{21}^2 = m_1^2 - m_2^2
= x(3d + c)  
\eqno(2.9)
$$
and
$$
\Delta m_{\rm atm}^2 = \Delta m_{23}^2 = m_2^2 - m_3^2
= (3(d+c)-x)(c-3d+x)/4 
\eqno(2.10)
$$
\noindent
The best fit values of oscillation
parameters from solar neutrino experiment
, atmospheric neutrino
experiments and solar neutrino mixing angle  
 are used to obtain the values of
$a$, $c$ and $d$.  
Before going to the neumerical fit of the
parameters, we would like to mention that the 
present mass matrix contains three parameters which are 
fitted with three experimental values, hence, the 
the texture has predictions about the neutrinoless 
double beta decay result and cosmological constraint on 
total neutrino mass (WMAP experimental result).  
The predictibility of the texture would increase 
if we incorporate more symmetries in an explicit 
model, which 
necessarilly reduce the number of  parameters. 
However, since the present texture admits nicely the constraint 
from WMAP experiment ,the only testability 
of the present texture relies crucially on the future 
experimental 
results of neutrinoless double beta decay. If this experimental 
result gives a strong evidence in favour of Majorana-type 
neutrino mass, then the present texture should explain 
the result or it will be ruled out.   
In the next section we will fit those 
values with the best fit results 
through our defined $\chi_p^2$ function. 
\section{Neutrino phenomenology}   

\noindent
The best fit values of oscillation
parameters from solar neutrino experiment
and atmospheric neutrino
experiments are used to obtain the values of
$a$, $c$ and $d$.  For this purpose, we consider                         
$\Delta m_{12}^2$, the
difference of the square
of mass eigenstates $m_1$ and $m_2$ and
the mixing angle $\theta_{12}$ that
are responsible for solar neutrino oscillation
and  $\Delta m_{23}^2$,
$\theta_{23}$ responsible for oscillation
of atmospheric neutrinos.
A recent global analysis \cite{sno}
of the solar neutrino data
from all solar neutrino experiments namely Chlorine, Gallium,
Super-Kamiokande, SNO charged current, neutral current including 
salt phase data, shows that large mixing angle or
LMA solution is most
favoured for solar neutrino oscillation. From the combined 
analysis with solar neutrino data and KamLand experiment
the best fit values of the solar neutrino oscillation parameters
are given by $\Delta m_\odot^2 (\equiv \Delta m_{12}^2$
for our model) $= 8 \times
10^{-5}$ eV$^2$ and tan$^2\theta_\odot
(\equiv$ tan$^2\theta_{12}$ for our model)
$= 0.45$ \cite{sno}. 
From the analysis of SK atmospheric neutrino oscillation data,
\cite{skbest} we have the best fit values of $\Delta m_{\rm atm}^2
(\equiv \Delta m_{23}^2$
for our model) $= 2.4 \times 10^{-3}$ eV$^2$ and $\theta_{\rm atm}$
is maximal. This value of $\theta_{\rm atm}$
has already been obtained in our proposed texture
for $\theta_{23}$. Thus treating $a$, $c$ and $d$ as
parameters we can obtain different values of
$\Delta m_{12}^2$, $\Delta m_{23}^2$ and tan$^2\theta_{12}$ and
compare them with best fit values of those quantities namely
$\Delta m_\odot^2$, $\Delta m_{\rm atm}^2$ and tan$^2\theta_\odot$
obtained from the analysis of solar and atmospheric neutrino 
data (discussed above)
to fix $a$, $c$ and $d$. To this end we define a function
$$
\chi^2_p = (1- \Delta m_{12}^2/\Delta m^2_\odot)^2
+ (1 - \Delta m_{23}^2/\Delta m^2_{\rm atm})^2  $$ $$
 +  (1 - \tan^2\theta_{12}/\tan^2\theta_\odot)^2\,\, .
\eqno(2.11)
$$
The function $\chi^2_p$ as defined above is calculated for                      a wide range of values of
$a$, $c$ and $d$ and the minimum of the function is obtained.
The corresponding values of $a$, $c$ and $d$ are given below.
$$
\,\,\,\,\,\,\,\,\,\,\,\,\,\,\,\,\,\,\, a \,\, = -0.\,\,145\,\,\,{\rm eV} 
$$
$$
\,\,\,\,\,\,\,\,\,\,\,\,\,\, c \,\,\,\, =  -0.\,\,152\,\,\, {\rm eV}
$$
$$
\,\,\,\,\,\,\,\,\,\,\,\,\,\,\,\,\,\,\,\,\,
\,\, d \,\,\,\,= 0.\,\,051\,\,\, {\rm eV}
\eqno(2.12)
$$
\noindent
$\Delta m_{12}^2$, $\Delta m_{23}^2$ and $\tan^2\theta_{12}$ obtained
from the above values of $a$, $c$ and $d$ and their comparison       
with the best fit values for
$\Delta m^2_\odot$, $\Delta m^2_{\rm atm}$ and $\tan^2\theta_{\odot}$
obtained from recent analysis of the solar and atmospheric neutrino data
are shown in Table 1.
\noindent
In order to find out the range
of values of $a$, $c$ and $d$ that satisfy the
1$\sigma$ limits of $\Delta m_\odot^2$
and $\tan^2\theta_\odot$ for the  
(\cite{sno})
combined analysis of recent
solar neutrino data and KamLand data,
we have fixed the value
of $\Delta m_{23}^2$ at its best fit value (shown in Table 1) 
and vary the parameters $a$, $c$ and $d$ such  that
$\Delta m^2_{12}$ and $\tan^2\theta_{12}$ satisfy the allowed
range mentioned above.
This range as given in Ref. \cite{sno}
is $7.6 \times 10^{-5} eV^2 < \Delta m_\odot^2 < 8.6 \times 10^{-5} eV^2$ and
$0.38 < \tan^2 \theta_\odot < 0.54$.
The allowed region in parameter
space of $a$, $c$ and $d$ are shown in a 3D plot (Fig. 1). 
As seen from Fig. 1, the allowed
region is confined in a very narrow band in 
the parameter space. We have also found the parameter space 
in $a$, $c$ and $d$ for which $\Delta m_{23}^2$ lies in the 90$\%$ c.l.
range given by $2.0 \times 10^{-3} eV^2 \leq \Delta m_{23}^2 \leq 
5.0 \times 10^{-3} eV^2$ while
the other oscillation parameters kept in the previous range. This range of 
$\Delta m_{23}^2$ with 90 \% c.l for SK 
atmospheric neutrino data is obtained from 
\cite{skbest}. The allowed region in the parameter 
space of $a$, $c$ and $d$ are shown in the 3D plot of 
Fig. 2.
The value of $d$ gives rise to the value of $M$ as 
$$
M = {\langle \phi \rangle}^2/d \simeq {10}^{15} {\rm GeV}.
\eqno(2.13)
$$
assuming $\langle \phi \rangle $ = 250 GeV.
This value of $M$ could be recognized as the right-handed symmetry 
breaking scale in some models 
which contains an extra $U(1)_R$ or $SU(2)_R$ 
symmetry in which neutrino masses are generated 
through see-saw mechanism.
Again, the sum of the three neutrino mass comes out as 
$\Sigma m_i $ = 0.44 eV which is far below from the upper 
bound of WMAP experimental result. 
Furthermore, the value of $d$ 
obtained from the best fit result 
is at the lower end of  the present limit on Majorana neutrino mass 
$(\langle m_{ee} \rangle = 0.05-0.84)$eV \cite{2beta}
obtained from the neutrinoless double beta decay experiment. Although the 
result is somewhat controversial \cite{contro}, however, the 
testability of the present model lies crucially on the future 
experiments such as, EXO, MOON, GENIUS etc.

\section{Conclusion}

We demonstrate  that quark-lepton complementarity 
can be realised in a 
'Bimaximal + democratic' type neutrino mass 
matrix  
which includes two possible sources of neutrino mass. 
One of them is the usual 'Bimaximal' type arise in different models 
and a possible source of 'Democratic' type is through the  
incorporation of higher dimensional mass terms assuming equlity 
between yukawa couplings. 
We then fix the  parameters through our defined 
$\chi_p^2$ function with the two mass squared 
differences solar and atmospheric and with the 
solar neutrino mixing angle.
The texture admits a best fit result at the minimum value 
of $\chi_p^2$. Furthermore, the texture gives rise to a slightly lower value 
of $m_{ee}$ mass compare to the experimental result of neutrinoless 
double beta decay experiment, and if the result survives in future with 
MOON, EXO, GENIUS experimental results, the present texture will be ruled 
out.    
\begin{acknowledgments}
Authors acknowledge Y. Koide 
for many helpful discussions and suggestions.
\end{acknowledgments}

\begin{table}
\begin{tabular}{|c|c|}
\hline
{\rm Present Work}& Experiment\\
\hline
$\Delta m^2_{12}$ & $\Delta m^2_{\odot}$ {\rm Ref.\cite{sno}}\\
(eV$^2$)          &      (eV$^2$)       \\
\hline
& \\
$8.02 \times 10^{-5}$ &
$8.0 \times 10^{-5}$ \\
& \\
\hline
$\Delta m^2_{23}$ & $\Delta m^2_{\rm atm} $ Ref.\cite{skbest} \\
(eV$^2$)          &      (eV$^2$)       \\
\hline
& \\
$2.6 \times 10^{-3}$ & $2.4 \times 10^{-3}$ \\               
& \\
\hline
$\tan^2\theta_{12}$ & $\tan^2\theta_\odot $ Ref.\cite{sno} \\
\hline
& \\
0.476 & 0.45 \\
& \\
\hline
\end{tabular}
\caption{
Best fitted values for          
neutrino oscillation parameters obtained
in the present work. 
}
\end{table} 
\newpage
\begin{center}
{\bf Figure Caption}
\end{center}
 
\noindent Fig. 1 The 3D plot showing the  region of the parameters
$a$, $c$ and $d$ that produce the values of $\Delta m_{12}^2$ and
$\tan^2\theta_{12}$ within 1$\sigma$ range of the best fitted
values from global solar neutrino + KamLand data analysis \cite{sno}.
$\Delta m_{23}^2$ remains fixed at the best fit value. 
See text for details. 

\vskip 1cm

\noindent Fig. 2 Same as Fig. 1 but in this case instead of 
keeping $\Delta m_{23}^2$ fixed at the best fit value, 
it is varied within the 90\% C.L. range
$2 \times 10^{-3} < \Delta m_{23}^2  < 5 \times 10^{-3} eV^2$
\cite{skbest}.

\begin{thebibliography}{99}
\bibitem{SK}
Super-Kamiokande Collaboration, Y. Fukuda et al., Phys.
Lett. B {\bf 433}, 9 (1998), B {\bf 436}, 33 (1998).
\bibitem{k2k}
E. Aliu et al (The K2K collaboration), Phys. Rev. Lett. {\bf 94}, 081802
(2005).
\bibitem{skbest}
G. Giacomelti, M. Giorgino, hep-ex/0504002.
\bibitem{sno}
B. Aharmin et al, (SNO collaboration), nucl-ex/0502021.
\bibitem{kam}
K. Eguchi et al, Phys. Rev. Lett. {\bf 90}, 021802 (2005), T. Araki 
et al, hep-ex/0406035
\bibitem{chooz}
T. Apollonio et al, Phys. Lett. B {\bf 466}, 415 (1999), {\it ibid}. 
B {\bf 420}, 397 (1998).
\bibitem{choozcom} G. Fogli, G. Lettera, E. Lisi, A. Palazzo and 
A. Rotunno, Phys. Rev. D {\bf 66}, 093008 (2002); M. Maltoni, 
T. Schwetz, M. Tortola and J.W.F. Valle, {\it ibid.} {\bf 67}, 
013011 (2003), M.C. Gonzalez-Garcia, M. Maltoni, C. Pena-Garay and 
J.W.F. Valle, Phys. Rev. D {\bf 63}, 033005 (2001).  
\bibitem{wmap}
C.L. Benett et al, Astrophys. J. Supl. {\bf 148}, 1 (2003).
\bibitem{wmapres}
V. Barger, D. Marfatia, A. Tregre, Phys. Lett. B {\bf 595}, 55 (2004).
\bibitem{2beta}
H. V. Klapdor-Kleingrothaus, A. Dietz, H. L. Harney,
and I. V. Krivosheina, Mod. Phys. Lett. {\bf A16},
2409 (2002).
\bibitem{qlc}W. Rodejohann, Phys. Rev. D {\bf 69}, 033005 (2004);
A. Y. Smirnov, hep-ph/0402264; S. F. Kang, C. S. Kim, J. Lee, 
hep-ph/0501029.

\bibitem{contro}
see for references H. V. Klapdor-Kleingrothaus, hep-ph/0205228 .
\end{thebibliography}
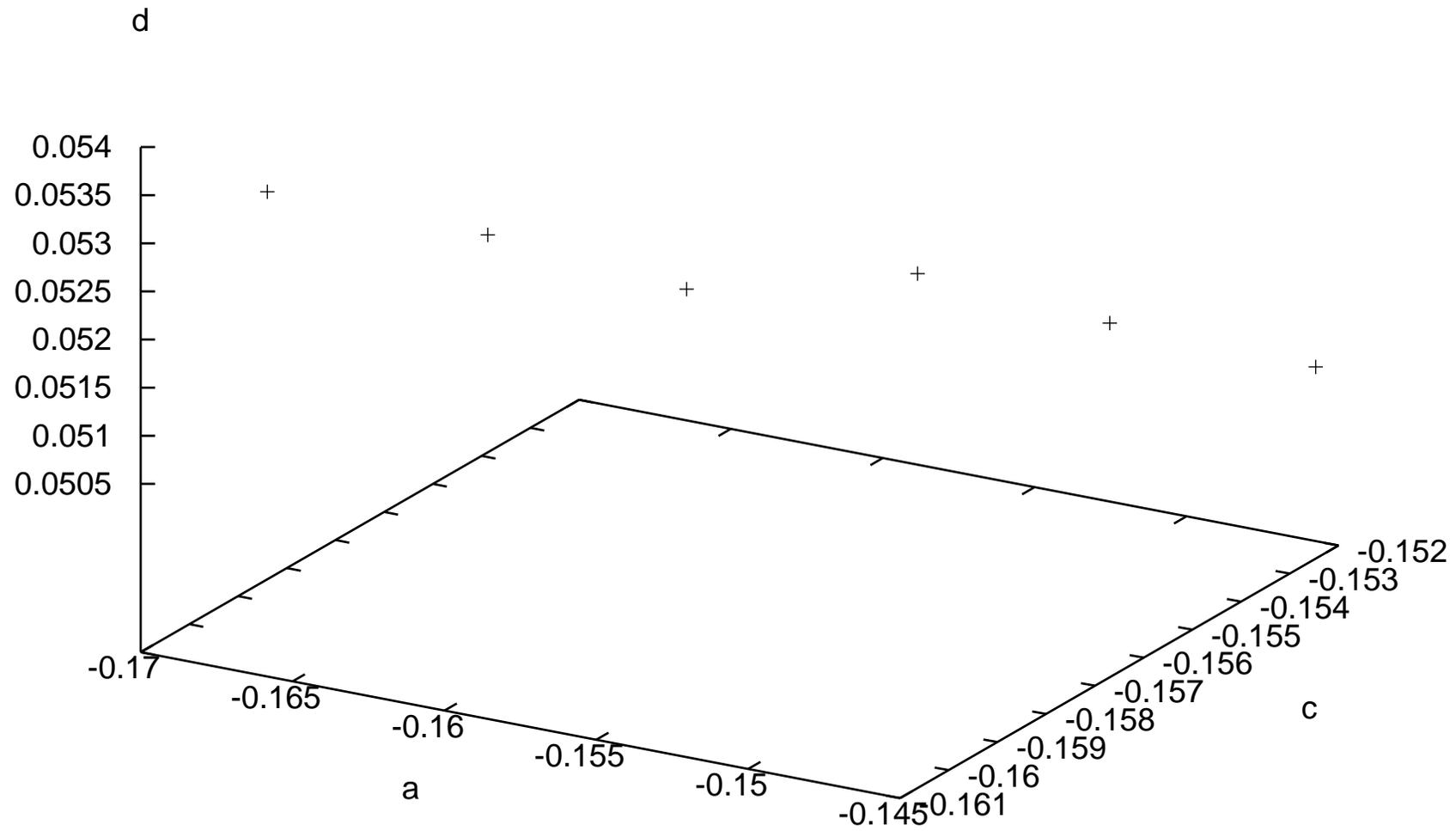
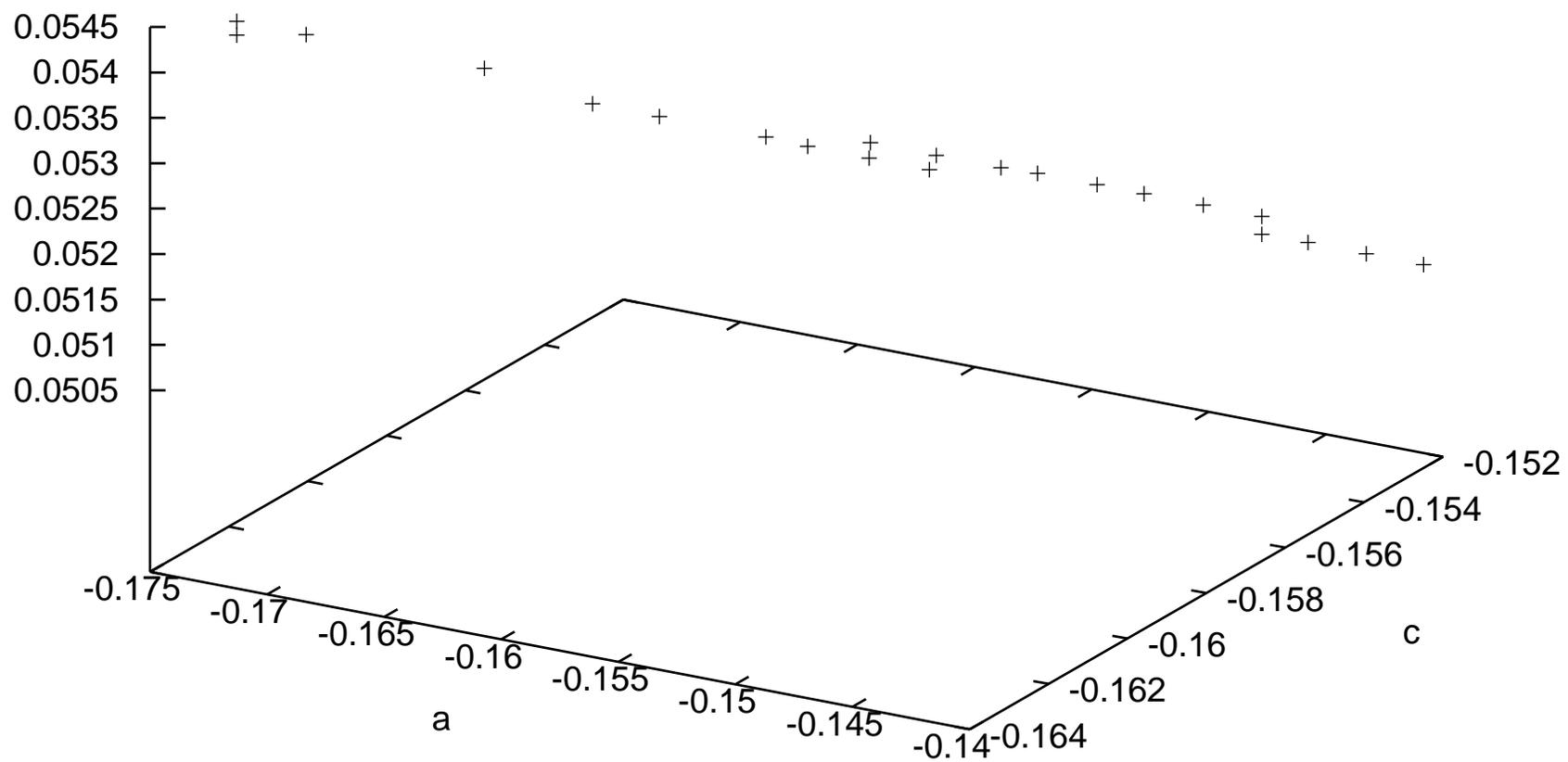
\end{document}